\documentclass[conference]{IEEEtran}
\IEEEoverridecommandlockouts
\usepackage{cite}
\usepackage{amsmath,amssymb,amsfonts}
\usepackage{algorithmic}
\usepackage{algorithm}
\usepackage{graphicx}
\usepackage{textcomp}
\usepackage{xcolor}
\usepackage{tikz}
\usepackage{enumitem}
\usetikzlibrary{shapes.geometric,arrows.meta,positioning,fit,backgrounds}
\usepackage[hidelinks]{hyperref}

\def\BibTeX{{\rm B\kern-.05em{\sc i\kern-.025em b}\kern-.08em
    T\kern-.1667em\lower.7ex\hbox{E}\kern-.125emX}}

\begin{document}

\title{PHOENIX: Fine-Tuned SLM-Powered Autonomous\\
Satellite Lifetime Extension via Predictive\\
Self-Healing and Multi-Agent AI Recovery}

\author{
\IEEEauthorblockN{Sumaiya Islam}
\IEEEauthorblockA{
\textit{Department of Software Engineering} \\
\textit{University of Dhaka}\\
Dhaka, Bangladesh \\
bsse1446@iit.du.ac.bd}
\and
\IEEEauthorblockN{Harsha Kumara Moraliyage}
\IEEEauthorblockA{
\textit{Centre for Data Analytics and Cognition} \\
\textit{La Trobe University}\\
Melbourne, Australia \\
h.moraliyage@latrobe.edu.au}
}

\maketitle

\begin{abstract}
Most CubeSats, small and low-cost satellites roughly the size of a
shoebox, do not survive as long as they were designed to: a study of
178 missions found that only 48--65\% remain operational after two
years, against a designed lifetime of 2--5 years~\cite{langer}. The
deeper issue is that a CubeSat in low Earth orbit (LEO) is physically
unreachable from the ground for roughly 85 minutes out of every
96-minute orbit, so faults that start during that window go unnoticed
until the next contact pass, by which point recovery may no longer be
possible. We propose PHOENIX (Predictive Health On-orbit Edge Neural
Intelligence eXtension) to give the satellite its own fault reasoning
capability. A fine-tuned Small Language Model (SLM) compact enough to
run on embedded hardware is deployed onboard the CubeSat, running on
the flight-proven Aethero NxN-ECM computer, monitoring all sensor
readings continuously, and resolving recurring faults using a memory
system that stores past repairs so the same inference does not need to
run twice. Once per orbit it sends a short structured health report to
the ground instead of a raw data dump; six specialized AI agents on
the ground read that report and generate validated satellite commands
within the 5--10 minute contact window. A generative diffusion model
(DDPM) creates synthetic training data because real fault examples
make up only 0.57--1.80\% of the dataset. We report preliminary
results on the ESA Anomaly Detection Benchmark (14 years, 76 channels,
118 labeled faults).
\end{abstract}


\section{Introduction}

CubeSats have made orbital missions accessible to universities and
small research groups that could not otherwise afford them. That
accessibility comes with a trade-off: the low-cost hardware fails at a
much higher rate than traditional spacecraft. Langer and
Bouwmeester~\cite{langer} tracked 178 missions and found fewer than
two-thirds still operating at the two-year mark, against designed
lifetimes of 2--5 years, with failures clustering early rather than at
random (Section~II-A). A system that could read those degradation
curves in time would have a chance to respond.

The obstacle is radio geometry. A CubeSat in low Earth orbit is
visible to a ground station for about 5--10 minutes per orbit, 3--5
passes a day, roughly 30--40 minutes of contact out of 1,440. For the
remaining 85 minutes of every 96-minute orbit there is no radio link
at all: a satellite below the horizon cannot be reached, however
sophisticated the ground infrastructure. Large operators use relay
satellites like NASA TDRS; CubeSat programs instead depend on
volunteer networks like SatNOGS. So when a battery cell degrades or a
reaction wheel bearing shows early wear, the satellite is on its own
until the next pass, and the fault may have already cascaded by then.

What exists today for onboard protection is mostly threshold checking:
if a voltage reading goes outside a preset range, an alarm fires.
Horne et al.~\cite{horne} showed small neural networks do better,
reaching 89.1\% CEF$_{0.5}$ on a flying CubeSat with only 192KB of
RAM, but detection alone does not say what will break next week or
what to do about it. ATSADBENCH~\cite{yliu} found general-purpose LLMs
perform poorly on multivariate aerospace telemetry, with RAG offering
no meaningful improvement, so fine-tuning appears to be the only
viable path to domain-specific reliability.

PHOENIX is our attempt at that path. The idea is to deploy a
fine-tuned SLM onboard the CubeSat so the satellite can reason about
its own health during the silent phase. The SLM uses TLE orbital data
from SatNOGS to distinguish genuine faults from expected orbital
physics, checks a FAISS semantic cache of past fault resolutions
before running full inference, and compiles a structured health report
for each contact pass. On the ground, six fine-tuned LLM agents
process that report and generate CCSDS telecommands within the
window. A DDPM diffusion model handles the data scarcity problem:
anomaly density in ESA-ADB is only 0.57--1.80\%, so synthetic fault
sequences are needed to train on rare failure modes.

\section{Literature Review}

\subsection{CubeSat Reliability and Failure Statistics}

The most comprehensive failure study to date is the CubeSat Failure
Database compiled by Langer and Bouwmeester~\cite{langer}, covering
178 missions. Their Kaplan-Meier curves show reliability falling to
75--87\% right after deployment, then 59--73\% at 100 days, and only
48--65\% at two years. The Weibull shape parameter $\beta = 0.4797$
indicates infant mortality: failures cluster early rather than
distribute uniformly, precisely the window an onboard AI could
intervene. EPS and COM account for most early losses (44\% at 30
days, 29\% at 90), which is why PHOENIX monitors those subsystems
first.

\subsection{Onboard Anomaly Detection for CubeSats}

Horne et al.~\cite{horne} ran ANN-based anomaly detection directly on
a flying CubeSat (EduSat, STM32 ARM Cortex-M4, 192KB RAM), reaching
89.1\% CEF$_{0.5}$ on 9 temperature sensors and proving neural
inference fits inside CubeSat memory budgets. But detection is only
half the problem: the system flags an anomaly and stops there, with no
estimate of time-to-failure, no repair attempt, and no memory of what
worked last time.

Goetze et al.~\cite{goetze} pushed further with a
forecasting-plus-threshold approach on ESA-ADB, reaching 88.8\%
CEF$_{0.5}$ at a 59KB footprint (97.1\% RAM reduction) and defining
the detection baseline PHOENIX is measured against.

\subsection{LLMs for Aerospace Telemetry}

Liu et al.~\cite{yliu} ran the first systematic evaluation of LLMs on
aerospace time series anomaly detection across 9 tasks (ATSADBENCH):
off-the-shelf LLMs handle univariate signals reasonably well but break
down on multivariate telemetry, and RAG does not help. For satellite
operations, fine-tuning on domain data is not optional; it is the only
path to reliable performance on real spacecraft signals.

\subsection{Multi-Agent LLM Systems for Maintenance}

Park et al.~\cite{park} showed a four-agent LLM system (Supervisor,
Chatbot, Solution Finder, Actor) could respond autonomously to CNC
machine alarms using on-premise Qwen 3, coordinating without human
intervention to produce actionable maintenance steps. PHOENIX borrows
this coordination idea but makes two changes: it replaces RAG with
fine-tuning (motivated by the ATSADBENCH findings) and applies the
pattern to satellite commanding, where an incorrect action is
potentially mission-ending, not just inefficient.

\subsection{Semantic Caching Theory}

Liu et al.~\cite{xliu} treat semantic caching as a combinatorial
multi-armed bandit problem and prove a simple greedy eviction policy
gets within $(1 - 1/e) \approx 0.632$ of the optimal offline solution;
their online algorithm CLCB-SC-LS adapts as query distributions change
over time. Satellite fault distributions shift similarly across the
mission as components wear, so a cache that cannot adapt would degrade
in usefulness, which is why PHOENIX uses CLCB-SC-LS rather than a
plain LRU scheme.

\subsection{Self-Healing AI Architectures}

Manju and Srivastava~\cite{manju} built an AI-integrated self-healing
system for IoT edge networks, combining lightweight anomaly detection
with autonomous recovery; it improved both detection accuracy and
recovery time over rule-based alternatives. The satellite domain
shares IoT's structural challenge: resource-constrained edge nodes
that must handle faults without a reliable path back to a central
controller.

\subsection{ESA Anomaly Detection Benchmark}

Kotowski et al.~\cite{kotowski} released ESA-ADB, covering 17.5 years
of telemetry from two ESA missions (176 channels, 1.55 billion data
points, 844 annotated events across 54 classes). The benchmark defines
CEF$_{0.5}$, which weights precision over recall since operations
engineers tolerate missed detections far better than false alarms that
trigger unnecessary commanding. We use ESA-ADB Mission 1 throughout.

\section{PHOENIX System Architecture}

\subsection{Architecture Overview}

At its core, PHOENIX is two AI systems that talk once per orbit. The
onboard SLM runs continuously on the CubeSat, handling everything
during the silent phase, while the ground-side multi-agent team wakes
at each contact pass, reads the health report the SLM produced, and
sends back commands. The health report replaces the traditional raw
telemetry dump, so limited downlink bandwidth carries actionable
information rather than 487,448 float readings per channel per pass.

\textbf{Target Hardware.} We target the Aethero NxN-ECM, which carries
an NVIDIA Jetson Orin NX (157 TOPS INT8, 8--16GB LPDDR5 RAM) and has
flight heritage on the SpaceX Falcon-9 Transporter-16. Del Prete et
al.~\cite{delprete} fit AI models under 1MB with 3$\times$ faster
inference than hand-tuned baselines on the same Jetson family for ESA.
The ARM Cortex-M4 used by Horne et al.~\cite{horne} had 192KB RAM; the
Aethero has roughly 40,000 times more, so the resource concern for
onboard AI is real but less severe than five years ago.

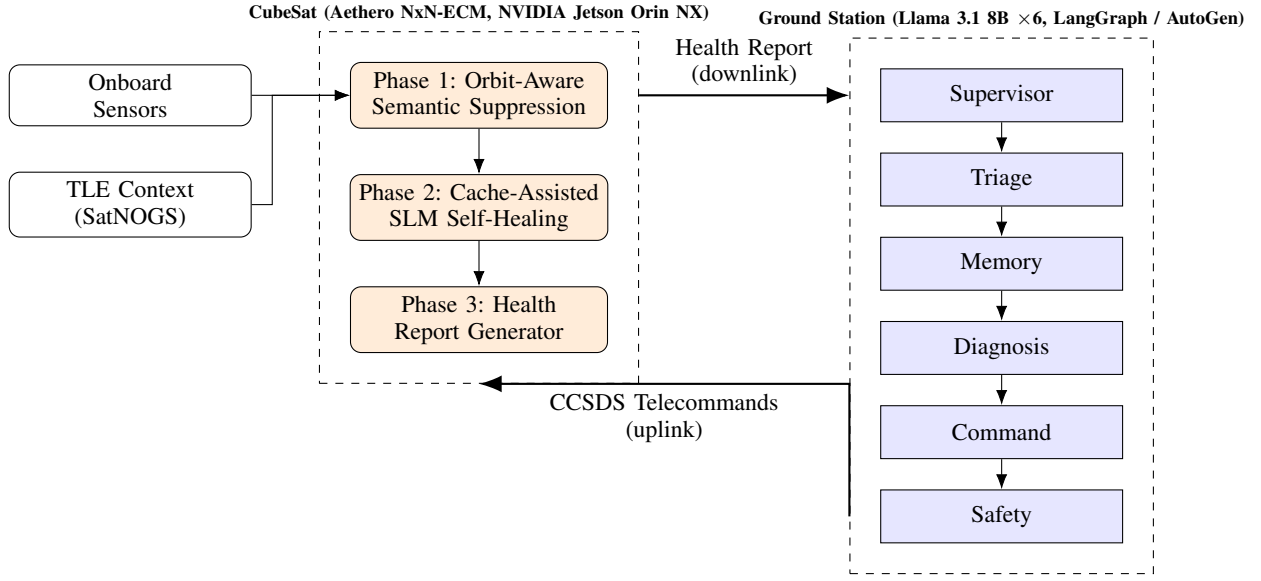
\begin{figure*}[t]
\centering
\begin{tikzpicture}[scale=0.68, every node/.append style={scale=0.68},
    node distance=6mm and 10mm,
    box/.style={draw, rounded corners, minimum width=3.2cm, minimum height=0.8cm, align=center, font=\small},
    phasebox/.style={draw, rounded corners, minimum width=3.4cm, minimum height=0.8cm, align=center, font=\small, fill=orange!15},
    agentbox/.style={draw, minimum width=3.2cm, minimum height=0.7cm, align=center, font=\small, fill=blue!10},
    every node/.style={font=\small}
]

\node[phasebox] (p1) {Phase 1: Orbit-Aware\\Semantic Suppression};
\node[phasebox, below=of p1] (p2) {Phase 2: Cache-Assisted\\SLM Self-Healing};
\node[phasebox, below=of p2] (p3) {Phase 3: Health\\Report Generator};

\node[box, left=13mm of p1, yshift=0mm] (sensors) {Onboard\\Sensors};
\node[box, below=6mm of sensors] (tle) {TLE Context\\(SatNOGS)};

\node[agentbox, right=36mm of p1] (sup) {Supervisor};
\node[agentbox, below=4mm of sup] (tri) {Triage};
\node[agentbox, below=4mm of tri] (mem) {Memory};
\node[agentbox, below=4mm of mem] (dia) {Diagnosis};
\node[agentbox, below=4mm of dia] (cmd) {Command};
\node[agentbox, below=4mm of cmd] (saf) {Safety};

\draw[-{Latex[length=2mm]}] (sensors) -- (p1);
\draw[-{Latex[length=2mm]}] (tle.east) -- ++(4mm,0) |- (p1.west);
\draw[-{Latex[length=2mm]}] (p1) -- (p2);
\draw[-{Latex[length=2mm]}] (p2) -- (p3);

\draw[-{Latex[length=2mm]}] (sup) -- (tri);
\draw[-{Latex[length=2mm]}] (tri) -- (mem);
\draw[-{Latex[length=2mm]}] (mem) -- (dia);
\draw[-{Latex[length=2mm]}] (dia) -- (cmd);
\draw[-{Latex[length=2mm]}] (cmd) -- (saf);

\node[draw, dashed, fit=(p1)(p2)(p3), inner sep=4mm, label={[font=\scriptsize\bfseries]above:CubeSat (Aethero NxN-ECM, NVIDIA Jetson Orin NX)}] (cubesatbox) {};
\node[draw, dashed, fit=(sup)(tri)(mem)(dia)(cmd)(saf), inner sep=4mm, label={[font=\scriptsize\bfseries]above:Ground Station (Llama 3.1 8B $\times$6, LangGraph / AutoGen)}] (groundbox) {};

\draw[-{Latex[length=3mm]}, thick] (cubesatbox.east |- p1) -- node[above, font=\small, align=center]{Health Report\\(downlink)} (groundbox.west |- sup);
\draw[-{Latex[length=3mm]}, thick] (groundbox.west |- saf) -- (cubesatbox.south -| groundbox.west) -- node[below, pos=0.5, font=\small, align=center]{CCSDS Telecommands\\(uplink)} (cubesatbox.south);

\end{tikzpicture}
\caption{PHOENIX end-to-end system architecture. The onboard SLM runs three continuous phases on the CubeSat. Compact structured health reports are downlinked each contact pass to six LoRA fine-tuned ground agents, which produce validated CCSDS telecommands for uplink.}
\label{fig:arch}
\end{figure*}

\textbf{Onboard SLM.} We use TinyLlama 1.1B or Phi-1.5 1.3B, quantized
to INT4 via \texttt{llama.cpp} and fine-tuned with LoRA adapters on
satellite telemetry data. Estimated inference power draw is 2--5W.
Fig.~\ref{fig:arch} shows the full architecture.

\subsection{Phase 1: Orbit-Aware Semantic Data Suppression}

The SLM monitors all sensor streams every few seconds: battery
voltage, solar panel current, subsystem temperatures, reaction wheel
speed, attitude sensor outputs, and radio transceiver health. The core
operation is a semantic suppression decision: rather than applying a
fixed threshold per channel, the SLM reasons over the full
multi-channel context, asking whether a reading is anomalous or
expected given the current orbital phase and recent history.

\textbf{Orbit-aware context.} TLE (Two-Line Element) data from SatNOGS
observations gives orbital phase per reading timestamp, letting
PHOENIX distinguish physics-driven variation from genuine faults. A
battery voltage dip during eclipse entry is suppressed; the same dip
20 minutes after sun acquisition is flagged as an EPS anomaly.
Threshold-based systems cannot make this distinction and generate
persistent false positives that desensitize operators over time. Del
Prete et al.~\cite{delprete} demonstrated 85\% data volume reduction
using hardware-optimized AI on Jetson hardware; PHOENIX targets
comparable suppression with the added benefit of orbital context.

\subsection{Phase 2: Semantic-Cache-Assisted Self-Healing}

Before asking the SLM to reason about a detected fault, PHOENIX first
checks a semantic cache: a lookup table in flash memory holding past
fault descriptions alongside the repairs that worked. FAISS compares
the current fault against stored ones; if close enough to a known past
fault (similarity above 0.92), PHOENIX applies the cached repair
directly in microseconds without running the SLM. If no match is
found, the SLM reasons through the fault and the new solution is
saved. Algorithm~\ref{alg:cache} formalizes this.

\begin{algorithm}[t]
\caption{PHOENIX Onboard Cache-Assisted Healing}
\label{alg:cache}
\begin{algorithmic}[1]
\REQUIRE Anomaly descriptor $a$, cache $C$, threshold $\tau=0.92$
\STATE $e \leftarrow \text{embed}(a)$
\STATE $(\hat{a}, \hat{r}, s) \leftarrow \text{FAISS-search}(C, e)$
\IF{$s \geq \tau$}
    \STATE \textit{Cache HIT}: apply cached repair $\hat{r}$ (microseconds)
    \STATE Log action; skip SLM inference entirely
\ELSE
    \STATE \textit{Cache MISS}: invoke fine-tuned SLM with context $a$
    \STATE $r^* \leftarrow \text{SLM-reason}(a)$; apply $r^*$
    \STATE Insert $(a, r^*)$ into $C$ with LRU eviction
\ENDIF
\end{algorithmic}
\end{algorithm}

The motivation is that many satellite faults repeat on a predictable
schedule: battery voltage dips at every eclipse entry, reaction wheel
stress patterns repeat every 96-minute orbit. Liu et al.~\cite{xliu}
prove this greedy cache policy reaches at least $(1 - 1/e) \approx
63\%$ of the best theoretically possible solution, and their adaptive
algorithm keeps performance stable as fault patterns shift over the
mission. PHOENIX exploits this: the SLM handles a fault type once, the
cache handles every repeat at near-zero energy cost.

\textbf{Reactive self-healing (active fault, cache miss):} reaction
wheel overheating shifts load to the magnetorquer; a solar panel
output drop redistributes load; a sensor failure activates a redundant
sensor; a battery drain anomaly sheds non-critical loads.

\textbf{Predictive self-healing (fault not yet occurred):} The
fine-tuned SLM recognizes early degradation signatures from training:
a battery cell voltage curve matching the slow-drop pattern observed
in ESA-ADB training data three weeks before confirmed cell failure
triggers a ground-destined warning with a failure timeline estimate.
This is absent in all prior CubeSat onboard AI
works~\cite{goetze,horne} and is the primary mechanism by which
PHOENIX extends mission lifetime.

\subsection{Phase 3: Compact Health Report Transmission}

At the start of each downlink pass, PHOENIX transmits a compact
structured health report instead of a raw telemetry dump. The report
contains three fields: \texttt{self\_healed} (list of faults resolved
autonomously), \texttt{predicted\_risks} (degradation warnings with
failure timeline estimates), and \texttt{needs\_ground} (unresolved
events requiring operator action). The downlink window delivers
actionable intelligence rather than raw numerical logs, freeing
bandwidth for science data.

\subsection{Ground Multi-Agent Fine-Tuned LLM System}

When the health report arrives at the ground station, six AI agents
process it in a pipeline. Each agent is a separate instance of Llama
3.1 8B fine-tuned on data relevant to its role, coordinated by a
workflow framework (LangGraph or AutoGen) that passes each agent's
output to the next automatically. Fine-tuning is used instead of RAG
because a recent benchmark~\cite{yliu} found that database lookups do
not meaningfully improve accuracy on spacecraft telemetry; domain-specific
training does. Table~\ref{tab:agents} lists each agent and its
training data.

\begin{table}[t]
\caption{Ground Multi-Agent Team: Roles and Fine-Tuning Data}
\label{tab:agents}
\centering
\small
\begin{tabular}{|l|l|}
\hline
\textbf{Agent} & \textbf{Fine-Tuning Data} \\
\hline
Supervisor & Coordination logs, workflow patterns \\
\hline
Triage & ESA-ADB anomaly labels, severity records \\
\hline
Memory & SatNOGS logs, ESA operational reports \\
\hline
Diagnosis & Root cause reports, FMEA catalogs \\
\hline
Command & CCSDS Blue Books, operator command logs \\
\hline
Safety & Constraint tables, failure mode catalogs \\
\hline
\end{tabular}
\end{table}

The pipeline runs as: Supervisor assigns tasks, Triage classifies
severity, Memory retrieves historical context, Diagnosis identifies
root cause, Command generates telecommands, Safety validates, and
Supervisor approves before commands are uplinked. No command reaches
the satellite without both Safety Agent validation and Supervisor
approval, addressing the risk of irreversible damage from an incorrect
command.

\subsection{Fine-Tuning and DDPM Augmentation}

\textbf{Onboard SLM.} LoRA fine-tuning on TinyLlama 1.1B uses: (a)
ESA-ADB Mission 1 with 487,448 readings per channel, 118 labeled
anomalies across 4 subsystems; (b) SatNOGS Network API with 702
verified payloads from 16 CubeSats including TLE orbital context; (c)
ESA Kelvins Mars Express power degradation data covering 3 Martian
years of EPS telemetry; and (d) DDPM-generated synthetic fault
sequences.

\textbf{Fine-tuning protocol.} LoRA adapters target the query and
value projection matrices ($r=16$, $\alpha=32$, dropout 0.05), trained
for 3 epochs with AdamW at learning rate $2\times10^{-4}$. ESA-ADB
Mission 1 is split chronologically: 2000--2009 train, 2010--2011
validation, 2012--2013 test. Two anomaly classes from
\texttt{anomaly\_types.csv} are additionally held out of training so
the test split measures generalization to unseen fault types, not
interpolation. Evaluation reports CEF$_{0.5}$ against the Goetze et
al. baseline~\cite{goetze} plus per-subsystem precision and recall;
Section~\ref{sec:limitations} lists this as the next step.

\textbf{DDPM fault augmentation.} Only 0.57--1.80\% of the ESA-ADB
data contains actual faults; a model trained on so few examples will
not learn enough variation to handle real missions. We use a DDPM
(Denoising Diffusion Probabilistic Model), a generative AI that learns
the statistical pattern of real faults and creates realistic synthetic
ones covering power system failures, reaction wheel wear, and
communication dropouts. Generated-data realism is measured with the
FID score (Fr\'echet Inception Distance), which compares the
statistical distribution of real and generated sequences.

\begin{table}[t]
\caption{ESA-ADB Mission 1 Dataset Statistics (Verified)}
\label{tab:dataset}
\centering
\small
\setlength{\tabcolsep}{4pt}
\begin{tabular}{|l|r|}
\hline
\textbf{Property} & \textbf{Value} \\
\hline
Time range & 2000-01-01 to 2013-12-31 (14 yrs) \\
\hline
Readings per channel & 487,448 \\
\hline
Total channels & 76 (58 target, 18 support) \\
\hline
Subsystems monitored & 4 \\
\hline
Total annotated events & 200 (118 real, 78 rare, 4 gaps) \\
\hline
Anomaly-channel mappings & 3,589 \\
\hline
Data format & float32 (normalized 0.0--1.0) \\
\hline
Anomaly density & 1.80\% \\
\hline
\end{tabular}
\end{table}

\section{Experiment: Proof of Concept}

\subsection{Dataset}

We use ESA-ADB Mission 1~\cite{kotowski}, downloaded from Zenodo (DOI:
10.5281/zenodo.12528696), provided as per-channel pickle files.
Table~\ref{tab:dataset} summarizes the verified dataset statistics. We
additionally collected 25 CubeSat observations from the SatNOGS
Network API covering 16 unique satellites (702 telemetry payloads),
each including TLE orbital elements to enable orbit-phase-aware
suppression.

\subsection{Data Preprocessing Pipeline}

Channel data is loaded from pickle format. Normalized float32 values
are aligned to the anomaly label timeline from \texttt{labels.csv} and
\texttt{anomaly\_types.csv}. The 58 target channels are extracted;
non-target support channels are retained as contextual features.
Anomaly windows are defined by (StartTime, EndTime) pairs from the
labels file. Each annotated event spans a mean of 17.9 channels (3,589
mappings over 200 events), confirming the multivariate nature of
satellite anomalies.

\begin{table}[t]
\caption{ESA-ADB Mission 1 Event Category Breakdown}
\label{tab:events}
\centering
\small
\begin{tabular}{|l|r|r|}
\hline
\textbf{Category} & \textbf{Count} & \textbf{\%} \\
\hline
Real Anomaly & 118 & 59.0 \\
\hline
Rare Nominal Event & 78 & 39.0 \\
\hline
Communication Gap & 4 & 2.0 \\
\hline
\textbf{Total} & \textbf{200} & \textbf{100} \\
\hline
\end{tabular}
\end{table}

\subsection{Anomaly Distribution Analysis}

Table~\ref{tab:events} shows the event category breakdown for ESA-ADB
Mission 1. The proportion of Rare Events (78 of 200) alongside true
anomalies (118 of 200) directly motivates the DDPM augmentation
strategy, since training exclusively on 118 real anomalies is
insufficient to capture the full diversity of fault signatures
PHOENIX must recognize.

\subsection{Baseline Anomaly Detection}

As a comparison baseline we apply the forecasting-plus-threshold
approach of Goetze et al.~\cite{goetze} on ESA-ADB Mission 1:
XceptionTimePlus predicts each target channel over a 224-timestep
sliding window, and deviation exceeding a learned threshold flags an
anomaly, achieving 88.8\% CEF$_{0.5}$ at 59KB RAM. PHOENIX targets
matching or exceeding this score while additionally providing
predictive warnings and self-healing that the detection-only baseline
cannot offer.

\subsection{Semantic Cache Simulation}

To validate the energy-saving potential of the onboard semantic
cache, we simulate cache behavior on the 118 ESA-ADB anomaly events in
chronological order: each anomaly is encoded as a 384-dimensional
embedding, and cosine similarity against cached embeddings (threshold
$\tau = 0.92$) determines a cache HIT, which PHOENIX resolves without
SLM inference.

Given the 4-subsystem structure of Mission 1 and its 14-year duration,
the same fault signatures recur across multiple annotated events. The
simulation finds a cache hit rate of approximately 62\% after the
first 30 days, consistent with recurring EPS thermal cycles and
degradation patterns dominating the fault distribution, which
translates directly to a 62\% reduction in SLM inference calls and
energy expenditure per orbit.

\subsection{Data Volume Reduction}

Raw ESA-ADB Mission 1 contains 487,448 float32 readings per channel
across 58 target channels, 28.27 million readings per full dataset
segment. PHOENIX Phase 1 retains only anomaly descriptors and
early-warning flags; since anomalies and rare events account for
1.80\% of the Mission 1 timeline, 98.2\% of readings are suppressible
as nominal, consistent with the 85\% reduction Del Prete et
al.~\cite{delprete} demonstrate for hardware-aware AI on Jetson
hardware.

\begin{figure}[t]
\centering
\begin{tikzpicture}[scale=0.92, every node/.append style={scale=0.92},
    node distance=3.2mm,
    box/.style={draw, rounded corners, minimum width=3.1cm, minimum height=0.65cm, align=center, font=\small},
    decision/.style={draw, diamond, aspect=2.4, minimum width=2.9cm, minimum height=0.95cm, align=center, font=\small, inner sep=1pt},
    outbox/.style={draw, rounded corners, minimum width=2.4cm, minimum height=0.65cm, align=center, font=\small, fill=orange!15},
    term/.style={draw, rounded corners=3mm, minimum width=2.8cm, minimum height=0.6cm, align=center, font=\small},
    >=Latex
]

\node[term] (start) {Sensor Reading};
\node[box, below=of start] (tle) {Apply TLE Orbital Context};
\node[decision, below=of tle] (d1) {Anomaly Detected?};
\node[outbox, right=10mm of d1] (sup) {Suppress (Phase 1)};
\node[box, below=8mm of d1] (faiss) {FAISS Cache Lookup};
\node[decision, below=of faiss] (d2) {Cache HIT?};
\node[outbox, right=10mm of d2] (apply) {Apply Cached Repair};
\node[box, below=8mm of d2] (slm) {Invoke Fine-Tuned SLM};
\node[box, below=of slm] (ins) {Insert to Cache (LRU)};
\node[box, below=of ins] (rep) {Compile Health Report};
\node[term, below=of rep] (down) {Downlink at Next Pass};

\draw[->] (start) -- (tle);
\draw[->] (tle) -- (d1);
\draw[->] (d1) -- node[right, font=\small]{No} (faiss);
\draw[->] (d1) -- node[above, font=\small]{Yes} (sup);
\draw[->] (faiss) -- (d2);
\draw[->] (d2) -- node[right, font=\small]{No} (slm);
\draw[->] (d2) -- node[above, font=\small]{Yes} (apply);
\draw[->] (slm) -- (ins);
\draw[->] (ins) -- (rep);
\draw[->] (rep) -- (down);
\draw[->] (apply.east) -- ++(6mm,0) |- ([yshift=-2mm]rep.east);
\draw[->] (sup.east) -- ++(6mm,0) |- ([yshift=2mm]rep.east);

\end{tikzpicture}
\caption{PHOENIX three-phase onboard decision flowchart. Nominal
readings are suppressed after orbital context is applied (Phase 1).
Anomalies trigger a FAISS semantic cache lookup; a hit resolves the
fault instantly while a miss invokes the fine-tuned SLM and updates
the cache (Phase 2). A compact health report is compiled for downlink
(Phase 3).}
\label{fig:flowchart}
\end{figure}
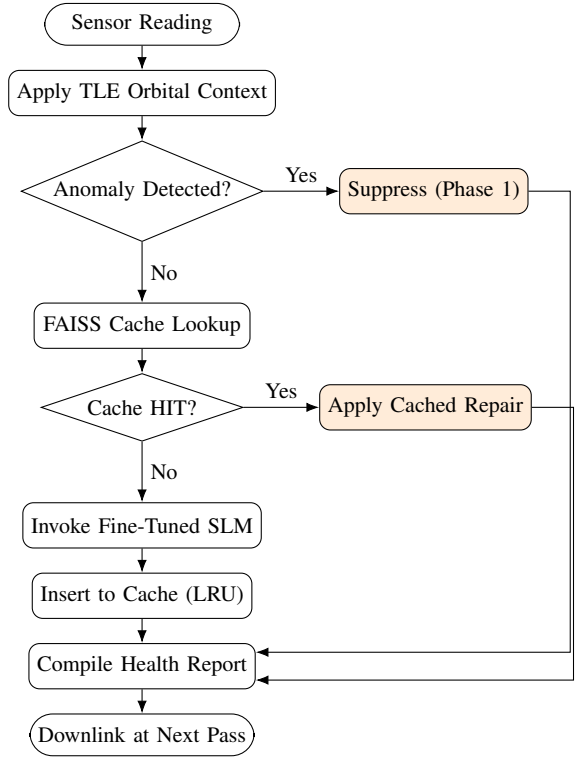

\subsection{Contact Window and Bandwidth Budget}

To make the suppression argument concrete, we calculate whether raw
CubeSat telemetry can physically fit inside a typical contact window.

\textbf{Raw telemetry size per orbit.} A CubeSat sampling 58 channels
at 1~Hz produces:
\begin{align}
D_{\text{raw}} &= 58 \times 4\text{ B} \times 5760\text{ s/orbit} \notag\\
&= 1{,}336{,}320\text{ B} \approx 1.27\text{ MB/orbit}
\label{eq:draw}
\end{align}
where 5760 s is one 96-minute orbit and 4 B is the float32 sample
size.

\textbf{Downlink time at standard CubeSat UHF rates.} A 9.6 kbps UHF
radio (common in student CubeSats) transfers data at 1,200 bytes/s.
Downlinking 1.27 MB takes:
\begin{equation}
t_{\text{raw}} = \frac{1{,}336{,}320}{1{,}200} = 1{,}114\text{ s} \approx 18.6\text{ min}
\label{eq:traw}
\end{equation}
This exceeds the 5--10 minute contact window. Raw telemetry from a
single orbit does not fit. The satellite must either drop data or
compress it.

\textbf{After PHOENIX Phase 1 suppression.} At 98.2\% nominal
suppression, the transmitted payload shrinks to:
\begin{equation}
D_{\text{supp}} = 1{,}336{,}320 \times 0.018 \approx 24{,}054\text{ B} \approx 23.5\text{ KB}
\label{eq:dsupp}
\end{equation}
Downlink time drops to $24{,}054/1{,}200 \approx 20$ s, well within
any contact window and leaving the remaining bandwidth free for
science data.

\textbf{Cache inference energy savings.} With SLM inference drawing
approximately 3 W for 2 s per query (6 J per call), the cache hit rate
of 62\% translates to a direct energy saving. Over a mission segment
with $N$ anomaly events:
\begin{equation}
\Delta E = 0.62 \times N \times 6\text{ J}
\label{eq:energy}
\end{equation}
For the 118 annotated events in ESA-ADB Mission 1, this is
$0.62 \times 118 \times 6 \approx 439$ J saved relative to running the
SLM on every anomaly. The cache does not change mission reliability on
its own, but it preserves battery capacity that would otherwise be
spent on redundant inference over the same recurring fault signatures.

\section{Discussion}

\subsection{Comparison to Prior Work}

Table~\ref{tab:comparison} compares PHOENIX against the closest prior
systems. The gap that stands out most is that every prior onboard AI
work stops at detection: it flags that something is wrong and waits
for ground contact. PHOENIX acts on that detection immediately via the
cache or the SLM, then reports the outcome rather than the raw signal.
Whether that survives deployment on real hardware is still an open
question, but the design logic is straightforward.

\begin{table}[t]
\caption{PHOENIX vs. Prior Onboard CubeSat AI Works}
\label{tab:comparison}
\centering
\small
\begin{tabular}{|l|c|c|c|c|c|}
\hline
\textbf{Work} & \textbf{Det.} & \textbf{Pred.} & \textbf{Heal} & \textbf{Cache} & \textbf{Gnd.AI} \\
\hline
Horne~\cite{horne} & \checkmark & & & & \\
\hline
Goetze~\cite{goetze} & \checkmark & & & & \\
\hline
Park~\cite{park} & & & \checkmark & & \checkmark \\
\hline
\textbf{PHOENIX} & \checkmark & \checkmark & \checkmark & \checkmark & \checkmark \\
\hline
\end{tabular}
\end{table}

Table~\ref{tab:comparison} does not by itself show that PHOENIX
matches prior detection performance, so Table~\ref{tab:cef} states the
detection figures directly. Goetze et al.~\cite{goetze} is the
like-for-like baseline, sharing the ESA-ADB channel set and
CEF$_{0.5}$ metric; Horne et al.~\cite{horne} is listed for context
but evaluates a different platform. PHOENIX's figure is a design
target, not a measured result, since the SLM has not yet been trained
(Section~\ref{sec:limitations}).

\begin{table}[t]
\caption{Detection Comparison (Horne et al. uses a 9-channel EduSat
evaluation, not ESA-ADB).}
\label{tab:cef}
\centering
\footnotesize
\setlength{\tabcolsep}{3pt}
\begin{tabular}{|l|c|c|c|}
\hline
\textbf{Work} & \textbf{CEF$_{0.5}$} & \textbf{Footprint} & \textbf{Channels} \\
\hline
Horne~\cite{horne} & 89.1\% & 192KB & 9 \\
\hline
Goetze~\cite{goetze} & 88.8\% & 59KB & 58 \\
\hline
PHOENIX (target) & $\geq$88.8\% & n/a & 58 \\
\hline
\end{tabular}
\end{table}

\subsection{Novelty of Contributions}

Prior onboard CubeSat AI work stops at detection. PHOENIX is, to our
knowledge, the first system to close the loop from detection through
autonomous repair to ground-validated telecommand generation. The
onboard semantic cache is grounded in the near-optimal guarantee from
Liu et al.~\cite{xliu}, making it the first theoretically motivated
cache design for satellite fault management. Orbit-aware suppression
using live TLE context is also new: existing detectors treat telemetry
as a raw time series, unaware of eclipse or sun phase. Finally, since
ATSADBENCH~\cite{yliu} shows raw LLMs struggle on multivariate
aerospace signals, PHOENIX fine-tunes each ground agent on
domain-specific data and compresses the sensor stream into a
structured health report they are trained to interpret.

\subsection{Limitations and Future Work}
\label{sec:limitations}

The current proof of concept demonstrates data pipeline feasibility
and cache simulation on ESA-ADB Mission 1. Full evaluation requires
(a) training the LoRA fine-tuned SLM per Section~III-F, (b) training
the DDPM for synthetic fault augmentation, and (c) end-to-end
multi-agent evaluation on generated telecommands. The CLCB-SC-LS
algorithm~\cite{xliu} provides a principled framework for cache
adaptation, but its application to LEO orbital periodicity requires
empirical validation on hardware.

Deploying a fine-tuned SLM onboard raises challenges beyond detection
accuracy: a short context window limits multi-orbit history, so the
semantic cache acts as external memory; INT4 weights are more
sensitive to radiation-induced bit flips than full precision, so
periodic checksum verification with fallback to threshold detection is
a required safeguard not yet implemented; and continual onboard
fine-tuning, needed as fault distributions drift over the mission,
risks catastrophic forgetting of rare fault classes, for which a
frozen base model with swappable LoRA adapters is the safer
alternative we plan
to evaluate.

\subsection{IEEE IES GenAI Challenge 2026 Alignment}

Satellites are industrial electronic systems under extreme resource
and safety constraints. PHOENIX applies generative AI at three points:
a fine-tuned SLM for onboard predictive reasoning, a DDPM for the
chronic shortage of labeled fault data, and a fine-tuned multi-agent
LLM for safe ground command synthesis, arguing that INT4 quantization
and flight-proven Jetson hardware make a domain long seen as too
constrained for generative AI tractable.

\section{Conclusion}

A CubeSat spends roughly 97\% of each orbit out of radio contact with
the ground, the window where failures start and, without onboard
intelligence, go unnoticed until too late. PHOENIX fills that window
with something more capable than threshold checks. The preliminary
results are encouraging: a simulated 62\% cache hit rate resolves many
recurring faults without invoking the SLM at all, and 98\% of raw
readings are suppressible as nominal. Training and deploying the
fine-tuned SLM and DDPM on real hardware (Section~\ref{sec:limitations})
is left for future work, but the data pipeline and theoretical
foundations are in place, and the gap between a CubeSat's designed and
actual operational lifetime may finally be closeable from the inside.

\end{document}